# The Hidden Nematic Fluctuations in the Triclinic $(Ca_{0.85}La_{0.15})_{10}(Pt_3As_8)(Fe_2As_2)_5$ Superconductor Revealed by Ultrafast Optical Spectroscopy

Qi-Yi Wu,[1] Chen Zhang,[1] Ze-Zhong Li,[2] Wen-Shan Hong,[2,3] Hao Liu,[1] Jiao-Jiao Song,[1] Yin-Zou Zhao,[1] Ya-Hua Yuan,[1] Bo Chen,[1] Xue-Qing Ye,[1] Shiliang Li,[2,4] Jun He,[1] H. Y. Liu,[5] Yu-Xia Duan,[1] Hui-Qian Luo,[2,4,*] and Jian-Qiao Meng[1,†]

[1]*School of Physics and Electronics, Central South University, Changsha 410083, Hunan, China*
[2]*Beijing National Laboratory for Condensed Matter Physics, Institute of Physics, Chinese Academy of Sciences, Beijing 100190, China*
[3]*International Center for Quantum Materials, School of Physics, Peking University, Beijing 100871, China*
[4]*Songshan Lake Materials Laboratory, Dongguan 523808, China*
[5]*Beijing Academy of Quantum Information Sciences, Beijing 100193, China*

We reported the quasiparticle relaxation dynamics of an optimally doped triclinic iron-based superconductor $(Ca_{0.85}La_{0.15})_{10}(Pt_3As_8)(Fe_2As_2)_5$ with bulk $T_c$ = 30 K using polarized ultrafast optical pump-probe spectroscopy. Our results reveal anisotropic transient reflectivity induced by nematic fluctuations develops below $T_{nem} \approx 120$ K and persists in the superconducting states. Measurements under high pump fluence reveal three distinct, coherent phonon modes at frequencies of 1.6, 3.5, and 4.7 THz, corresponding to $A_{1g}(1)$, $E_g$, and $A_{1g}(2)$ modes, respectively. The high-frequency $A_{1g}(2)$ mode corresponds to the $c$-axis polarized vibrations of FeAs planes with a nominal electron-phonon coupling constant $\lambda_{A_{1g}(2)} \approx 0.139 \pm 0.02$. Our findings suggest that the superconductivity and nematic state are compatible but competitive at low temperatures, and the $A_{1g}$ phonons play an important role in the formation of Cooper pairs in $(Ca_{0.85}La_{0.15})_{10}(Pt_3As_8)(Fe_2As_2)_5$.

The discovery of iron-based superconductors (FeSCs) has sparked a new upsurge in the study of high-temperature superconductors (HTSCs) [1]. Although the phase diagram of FeSCs is similar to that of cuprates, the details are different between both families. In particular, the normal state in cuprates is the well-known pseudogap, while its existence in FeSCs remains an open issue [2–4]. Recently, it has been reported that the nematic order is found in the normal state of FeSCs [5–7] and cuprates [8, 9], providing a new perspective for understanding the property of pseudogap. Since the nematic order breaks the fourfold rotational ($C4$) symmetry but preserves the translational symmetry, it presents as the in-plane anisotropy in various physical quantities. Elucidating the origin of nematic order is considered crucial to understanding the superconductivity in HTSCs. It has been theorized that the spin-driven nematic order and superconductivity with anisotropy gap symmetry ($s^{+-}$- and $d$-wave) are the consequence of the magnetic ground state, while the orbital-driven nematic order and conventional $s$-wave superconductivity are the results of the charging scenario [10–12].

Since the spin and orbital degrees of freedom are strongly coupled to each other, the challenge is how to distinguish experimentally the origin of nematic order [13–16]. Ultrafast spectroscopy offers the possibility to detect and clarify the nematic order in the time domain, based on the distinct relaxation dynamics of various orders [17–20]. Although it is widely accepted that the nematic order may originate from electrons rather than lattice distortion, recent Raman experiments have shown the presence of significant phonon anomalies in the nematic phase [21, 22]. Pump-probe technology can provide some information about lattice, as the coherent phonon oscillations with Raman activity can be excited by ultrafast lasers [23–26].

In comparison with the other iron-based superconductors, the recently discovered electron-doped $(Ca_{1-x}La_x)_{10}(Pt_3As_8)(Fe_2As_2)_5$ compound is more similar to the cuprate, whose superconductivity is achieved by charge doping of FeAs layers rather than the elementary substitution inside iron layers [27]. Its parent compounds $Ca_{10}(Pt_3As_8)(Fe_2As_2)_5$ (so-called 10-3-8 system) has a triclinic crystal structure and quasi-two-dimensional (quasi-2D) characters, and each iron arsenide (FeAs) layers have large separation ($c_0$ = 10.64 $Å$) caused by the intercalation of platinum arsenide ($Pt_3As_8$) layers and calcium (Ca) planes [Fig. 1(a)] [27, 28]. The parent compound shows a structural phase transition at $T_s$ = 103 K $\sim$ 110 K, followed by an antiferromagnetic (AFM) transition at $T_N$ = 90 $\sim$ 100 K [29–31]. For optimally doped $(Ca_{0.85}La_{0.15})_{10}(Pt_3As_8)(Fe_2As_2)_5$ (CaLa-10-3-8), the optical [32] and nuclear magnetic resonance measurements [33] indicate that it is a multiple-gap superconductor and has pseudogap behavior.

In this letter, we performed the polarized ultrafast optical pump-probe spectroscopy on the optimally electron-doped CaLa-10-3-8 superconductor with bulk $T_c$ = 30 K. Although the signs of the superconducting component along $a$ and $b$ axes are opposite below $T_c$, the extracted superconducting gaps using the Rothwarf-Taylor (RT) model are equal, i.e., $\Delta^0_{SC}(0) = \Delta^{90}_{SC}(0) \approx 11.5 \pm 0.1$ meV. In addition, we observed an anisotropic transient reflectivity induced by the nematic fluctuations that developed from $T_{nem} \approx 120$ K and continued below $T_c$. Three coherent phonon oscillations $A_{1g}(1)$, $E_g$, and $A_{1g}(2)$ were identified under a high pump fluence. The electron-phonon ($e$-$ph$) coupling constant $\lambda_{A_{1g}(2)}$ = 0.139 $\pm$ 0.02 is estimated for the highest frequency $A_{1g}(2)$ mode. Our results suggest that the nematic order is intertwined with superconductivity and coupled with the coherent phonon in CaLa-10-3-8.

In this study, high-quality single crystals of CaLa-10-3-8 with good cleavage planes (001) were grown using the self-flux method [33]. Ultrafast laser pulses of 800 nm ($\sim$1.55 eV) central wavelength, 35 fs pulse width, and 1 MHz repetition rate was used to pump and probe the ultrafast dynamics of CaLa-10-3-8 sample from 5 K to 300 K [24–26]. The polarization of the two beams was varied by rotating the half-wave plate and polarizer in front of the sample. The polarization angle $\phi$ is defined as the rotated angle relative to the horizontal polarization, as shown in Fig. 1(b). Measurements were performed on a freshly-cleaved surface under a $10^{-6}$ mbar vacuum.

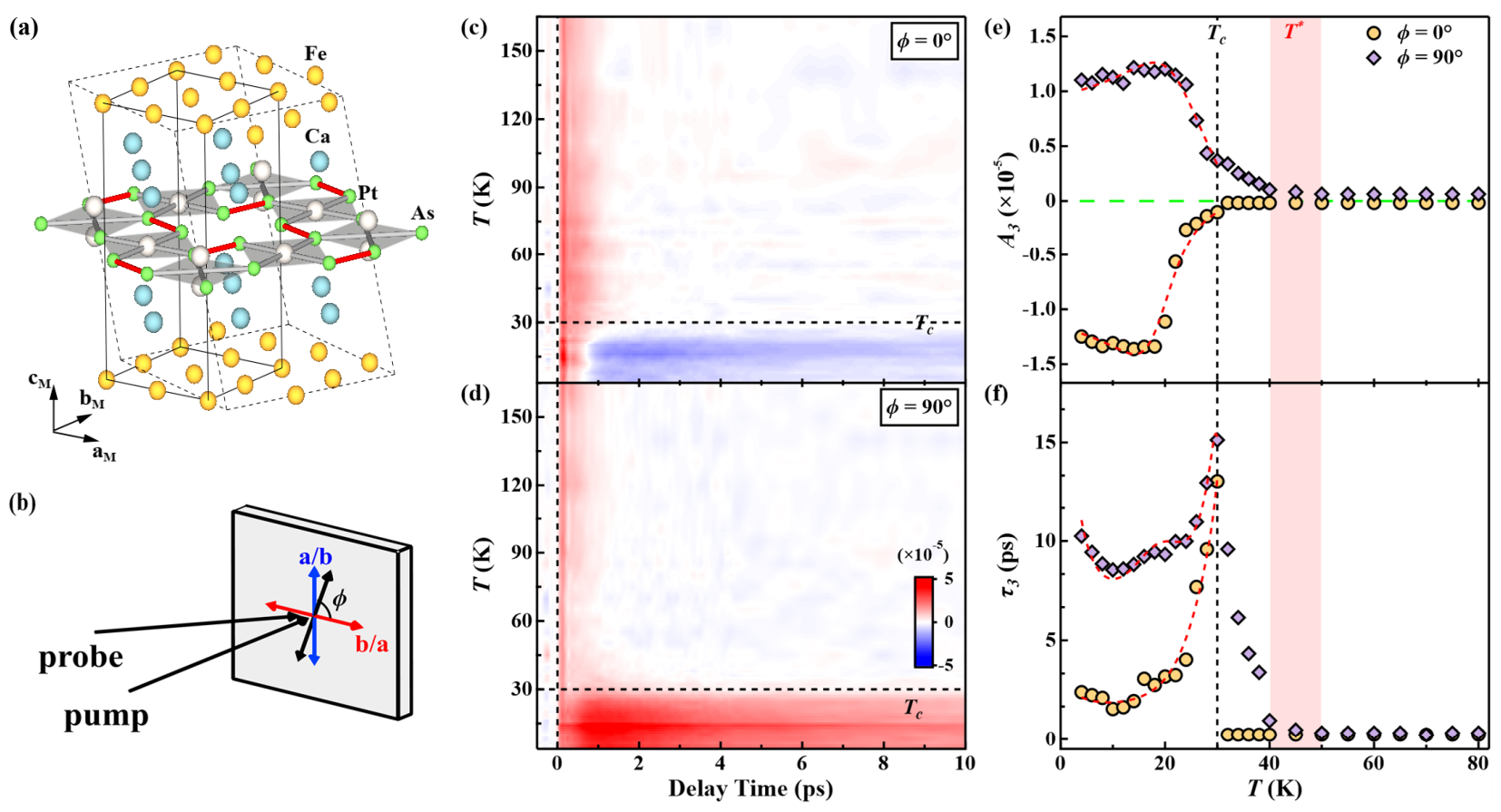

FIG. 1. (color online) (**a**) Crystalline structure of $(Ca_{0.85}La_{0.15})_{10}(Pt_3As_8)(Fe_2As_2)_5$. (**b**) The schematic diagram of the polarized pump-probe spectroscopy. (**c**) and (**d**) 2D color map of $\Delta R(t)/R$ as a function of temperature and delay time at a pump fluence of $\sim$3.8 $\mu$J/cm$^2$, probed with polarization angles $\phi$ = 0° and $\phi$ = 90°, respectively. (**e**) and (**f**) $T$-dependence of amplitudes $A_3$ and relaxation time $\tau_3$, respectively. The red dash lines are the RT model fitting curves.

Figures 1(c) and 1(d) present the 2D pseudocolor $\Delta R/R$ mapping images with temperature versus delay time along $\phi$ = 0° and $\phi$ = 90°, respectively. The relaxation process can be well fitted with the tri-exponential decays convoluted with a Gaussian laser pulse (see Fig. S2 of Supplemental Materials [34]):

$$\frac{R(t)}{R} = \frac{1}{\sqrt{2\pi}w}\exp(-\frac{t^2}{2w^2})\otimes[\sum_{i=1}^{3}A_i\exp(-\frac{t-t_0}{\tau_i})]+C, \quad (1)$$

where $A_i$ and $\tau_i$ represent the amplitude and relaxation time of the $i$th decay process, respectively. $w$ is the incidence pulse temporal duration, and $C$ is a constant offset. The initial and briefest relaxation process is always considered as an $e$-$e$ scattering process since its lifetime $\tau_1$ is comparable to the instrument's temporal resolution [24, 38].

Figures 1(e) and 1(f) summarize the temperature dependence of the SC response characterized by $A_3$ and $\tau_3$ under different polarizations ($\phi$ = 0° and $\phi$ = 90°). For both polarizations, the amplitude $A_3$ changes suddenly below $T_c$. The relaxation time $\tau_3$ divergent near $T_c$. These features indicate the opening of the superconducting gap, which can be well described by the RT model [24, 39–41],

$$A(T) \propto \frac{\varepsilon_I/[\Delta(T)+k_BT/2]}{1+\gamma\sqrt{2k_BT/\pi\Delta(T)}e^{-\Delta(T)/k_BT}}, \quad (2)$$

$$\tau \propto \frac{\hbar\omega^2\ln\{1/[\varepsilon_I/\alpha\Delta_{SC}(0)^2+e^{-\Delta(T)/k_BT}]\}}{12\Gamma_\omega\Delta(T)^2}. \quad (3)$$

where $\varepsilon_I$ is the absorbed laser energy density per unit cell. $\gamma$, $\omega$, $\alpha$, and $\Gamma_\omega$ are the fitting parameters. Figures 1(e) and 1(f) present a good fit to the experimental data below $T_c$, giving a zero-temperature gap of $\Delta_{SC}(0) \approx 11.5$ meV for both polarizations, with a typical Bardeen-Cooper-Schrieffer (BCS) form of temperature dependence gap $\Delta(T) = \Delta_{SC}(0)[1\text{-}(T/T_c)]^{1/2}$ [42, 43]. The SC gap obtained in our study is consistent with the larger SC gap obtained by optical conductivity measurements ($\approx$ 14.2 meV) [32]. It is noteworthy that even though $A_3$ exhibits obvious anisotropy as the polarization direction is rotated from $\phi$ = 0° to $\phi$ = 90°, it cannot give any information about the symmetry of the superconducting gap, because it can be attributed to the anisotropy of probe transition matrix elements or the anisotropy of the spectral weight transfer in the real part of optical conductivity [35, 44]. In addition, $A_{3,90}$ and $\tau_{3,90}$ fluctuate above $T_c$ and vanish around 45 K. This temperature is consistent with the characteristic temperature $T^*$ = 45 K obtained from recent inelastic neutron scattering (INS), Nernst effect, and nuclear magnetic resonance (NMR) experiments, which is related to the SC fluctuation [33]. Thus, $T^*$ could mark the onset of a pseudogap, which is likely associated with the emergence of preformed Cooper pairs [45].

Figure 2(a) compares the transient reflectivity along $\phi$ = 0° and $\phi$ = 90° at different temperatures. At high temperatures, the transient reflectivity along both polarizations is the same. As the temperature decreases, $\Delta R/R$ of the two polarizations begins to differ. Initially, the difference appears below $\sim$ 2 ps, and as the temperature further drops, the relative intensities of the two flip and the time range of the difference expands. The difference plot $(\Delta R/R)_{ani}$ = $(\Delta R/R)_{90}$ - $(\Delta R/R)_0$ is shown in Fig. 2(b) and reveals a clear polarization anisotropy. In addition to the significant positive $(\Delta R/R)_{ani}$ in the superconducting state, a negative $(\Delta R/R)_{ani}$, contributed by a sec-ond relaxation process, occurs below $\sim$ 120 K with a lifetime within 2 ps, which is suppressed below $T^*$ [the green region in Fig. 2(b)]. Figures 2(c) and 2(d) summarized the amplitudes and relaxation times of the second relaxation process along both polarizations. At high temperatures, the amplitude and relaxation time along both polarizations are almost equal and independent of the temperature. As the temperature cools below $\approx$ 120 K, both amplitudes and relaxation times along different polarizations show an opposite tendency with decreasing temperature, which implies a break of symmetry. This anisotropy is more clearly in the $A_{ani} = A_{2,0} - A_{2,90}$ and $\tau_{ani} = \tau_{2,0} - \tau_{2,90}$ [Figs. 2(e) and 2(f)]. Below $\approx$ 120 K, $A_{ani}$ shows an apparent reduction, and the $\tau_{ani}$ increases with the decreasing temperature. Therefore, it is reasonable to argue that here in $(Ca_{0.85}La_{0.15})_{10}(Pt_3As_8)(Fe_2As_2)_5$ the nematic fluctuations with an onset temperature at $T_{nem} \approx$ 120 K.

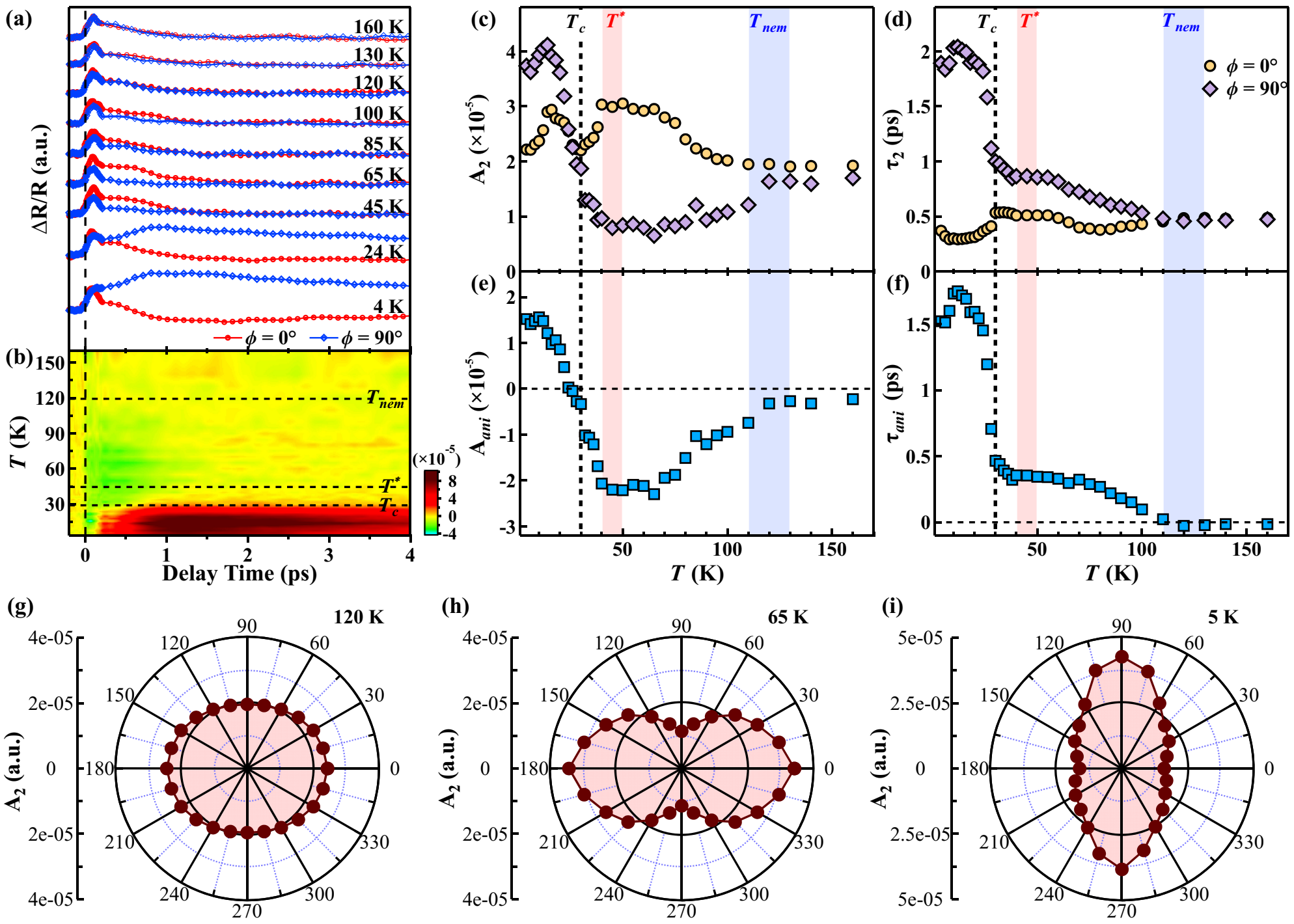

FIG. 2. (color online)(**a**) $\Delta R(t)/R$ as a function of delay time probed with $\phi = 0°$ and $\phi = 90°$ at selected temperatures. (**b**) The intensity difference between $\phi = 90°$ and $\phi = 0°$ [$(\Delta R/R)_{90}$ - $(\Delta R/R)_0$]. (**c**) and (**d**) $T$-dependence of amplitudes $A_2$ and relaxation time $\tau_2$, respectively. (**e**) and (**f**) Difference of amplitude $A_2$ and relaxation time $\tau_2$ between two polarization: $A_{ani} = A_{2,90} - A_{2,0}$, $\tau_{ani} = \tau_{2,90} - \tau_{2,0}$. (**g**) - (**i**) Polarization dependence of amplitude $A_2$ at selected temperatures.

To further elucidate the broken symmetry, we performed the polarized experiment at the selected temperatures and extracted the angular dependence of $A_2$ by using Eq. (1), as shown in Fig. 2 (g) - 2 (i). At a temperature of 120 K, approximately $T_{nem}$, $A_2$ is almost isotropic [Fig. 2(g)]. At 65 K, well below $T_{nem}$, $A_2$ at $\phi = 0°$ shows a maximum, while that at $\phi = 90°$ shows a minimum [Fig. 2 (h)]. $A_2$ shows twofold symmetry, clearly demonstrating the breaking of the fourfold rotational symmetry. This transient anisotropy is also observed in other FeSCs [17–20, 46] and attributed to the nematic fluctuations contributed by the vestigial phase from rotation-symmetry-broken electronic states. In our study, the orbital scenario can qualitatively explain the difference between $\phi = 0°$ and $\phi = 90°$. The existence of nematic order will cause the populations in $d_{xz}$ and $d_{yz}$ orbits to be different. Since the probe pulses polarized along the $a$ and $b$ axes are sensitive to $d_{xz}$ and $d_{yz}$ orbits, respectively, the anisotropy in the reflectivity can be understood as the consequence of the different populations of excited quasiparticles in $d_{xz}$ and $d_{yz}$ orbits. However, a spin origin of the nematic order is also possible because recent nuclear magnetic resonance (NMR) experiments found that the spin-lattice relaxation of CaLa-10-3-8 increases rapidly when warming above 150 K, which is probably related to a symmetry breaking in the spin degree of freedom [33]. It is necessary to investigate the ellipticity of the probe beam to confirm whether the nematic state is driven by spin fluctuations in the future.

Moreover, the facts that the $A_{ani}$ and $\tau_{ani}$ change simultaneously around $T^* = 45$ K. Near $T_c$, the relative amplitudes of $A_{2,0}$ and $A_{2,90}$ are reversed [Fig. 2(c)], resulting in a change in the sign of $A_{ani}$ [Fig. 2(e)]. As shown in Fig. 2(i), at 5 K, well below $T_c$, $A_2$ at $\phi = 90°$ shows a maximum, while that at $\phi = 0°$ shows a minimum. That is, the superconducting transition leads to a 90° rotation of the long axis of $A_2$, indicating that the nematic state is strongly coupled and competes with the superconductivity. When $T_{nem} > T > T_c$, the relax-ation processes are dominated by nematic fluctuations. After pump pulses weaken the nematic fluctuations, it will rapidly recover to equilibrium. However, below $T_c$, the relaxation processes are the combination of the nematic response and superconducting response. The pump pulses not only weaken the nematic fluctuations but also suppress the superconductivity. The non-equilibrium nematic order will not recover towards its equilibrium, but towards a time-dependent quasi-equilibrium with the form of superconducting order parameters. Thus, the nematic fluctuations below $T_c$ will be enhanced by the mutual repulsion and also affected by superconductivity [46]. According to this picture, the rotation of the long axis below $T_c$ can be naturally interpreted as: when temperature cross $T_c$, the nematic signals along $\phi$ = 90° will be enhanced by the positive superconducting response, while the nematic signals along $\phi$ = 0° will be reduced by the negative superconducting signal.

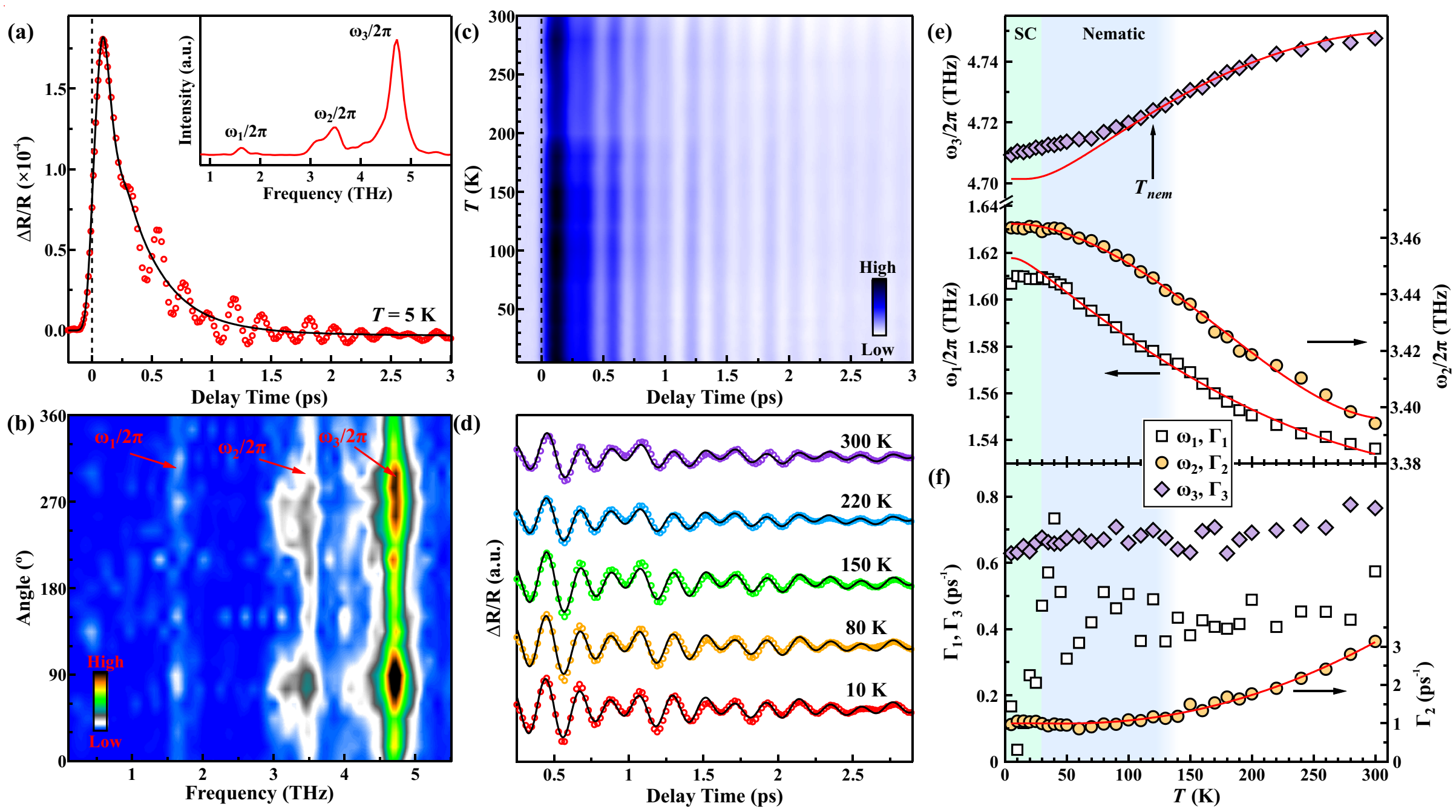

FIG. 3. (color online) (**a**) Typical $\Delta R(t)/R$ curves at a pump fluence of $\sim$ 122.8 $\mu$J/cm$^2$ along $\phi$ = 90°, composing of quasiparticle relaxation (black solid line) and coherent optical phonon vibration. Inset: FFT results showing three phonon modes with frequency of 1.6, 3.5, and 4.7 THz. (**b**) 2D map of FFT results as a function of polar angle and frequency at 5 K. (**c**) 2D color map of $\Delta R(t)/R$ as a function of temperature and delay time. (**d**) Phonon oscillations extracted at five temperatures. The black lines are Eq.(4) fits. (**e**) and (**f**) The derived frequency and damping rate, respectively, as a function of temperature. The red solid lines are fitted curves using the anharmonic phonon model.

Besides the quasiparticle relaxation dynamics under low fluence, we also focus on the behaviors of coherent phonon which is excited by the high fluence laser pulses. Apparent oscillations appear in $\Delta R(t)/R$, as shown in Fig. 3(a). The oscillatory components are obtained by performing the Fast Fourier transform (FFT) on the oscillation extracted by subtracting the exponential decay from the $\Delta R(t)/R$. The inset in Fig. 3(a) presents three terahertz modes with frequencies of $\omega_1/2\pi \sim$ 1.6 THz (i.e., 6.6 meV or 53.4 cm$^{-1}$), $\omega_2/2\pi$ $\sim$ 3.5THz (i.e., 14.5 meV or 116.7 cm$^{-1}$), and $\omega_3/2\pi \sim$ 4.7 THz (i.e., 19.4 meV or 156.8 cm$^{-1}$). To date, there has not been any Raman spectroscopy studies or phonon calculation of the Ca10-3-8 materials. However, other iron-based superconductor families with FeAs layers show Raman-active phonons near these energies. Both the lowest component $\omega_1$ and the primary component $\omega_3$ are $A_{1g}$ modes, labeled $A_{1g}(1)$ and $A_{1g}(2)$, respectively, corresponding to $c$-axis polarized vibrations of FeAs layers [47]. The $\omega_2$ shows prominent polarization-dependence [Fig. 3(b)], suggesting that this mode is the $E_g$ mode, which involves the in-plane asymmetry of FeAs layers [36, 48–50].

The ultrafast spectroscopy measurements on other FeSCs revealed that the *e-ph* coupling constant $\lambda_{A_{1g}}$ is proportional to the $T_c$ [37], suggesting that the $A_{1g}$ mode plays an essential role in forming Cooper pairs [37]. Therefore, we estimated the $\lambda_{A_{1g}(2)}$ of CaLa-10-3-8 from our data, i.e., $\lambda_{A_{1g}(2)} \approx 0.14 \pm 0.02$ (see calculation details in the Supplemental Material [34]). The $\lambda_{A_{1g}(2)}$ data point is located on the curves described in [37] within the error range, indicating that the superconductivity mechanism of CaLa-10-3-8 might have commonality with the other FeSCs, although the doping scenario of CaLa-10-3-8 is more similar to the cuprate in previous works [27].

Figure 3(c) presents the 2D pseudocolor mapping image of $\Delta R(t)/R$ as a function of temperatures and delay time. The oscillation survives up to room temperature and can be fitted by the following expression [black curves in Fig. 3(d)]:

$$(\frac{\Delta R}{R})_{osc} = \sum_{j=1}^{3} A_j e^{-\Gamma_j t} \sin(\omega_j t + \phi_j), \quad (4)$$

where $A_j$, $\Gamma_j$, $\omega_j$, and $\phi_j$ are the $j$th oscillatory signal amplitude, damping rate, frequency, and initial phase, respectively. The extracted temperature evolutions of frequencies and damping rates are plotted in Fig. 3(e) and 3(f). The temperature dependence of $E_g$, $A_{1g}(1)$ and $A_{1g}(2)$ phonons at high temperatures can be well described by the anharmonic effect [24, 51, 52]. However, it fails to capture the behaviors of $A_{1g}(1)$ and $A_{1g}(2)$ phonons at low temperatures. Below $T_{nem}$, the $\omega_3$ deviates from the fitting curve obtained by using the anharmonic effect with a negative thermal expansion coefficient, implying the $A_{1g}(2)$ phonon might couple with the nematic order [diamond in Fig. 3(e)]. Moreover, the phonon softening in $\omega_1$ and the divergence in $\Gamma_1$ were observed simultaneously around $T^*$ [empty square in Fig. 3(e) and 3(f)], indicating the $A_{1g}(1)$ mode might have a potential intimacy with the opening of superconducting gap $\Delta_{SC}$ [42].

In conclusion, we've used polarized ultrafast optical spectroscopy to investigate the quasiparticle dynamics and coherent phonons of $(Ca_{0.85}La_{0.15})_{10}(Pt_3As_8)(Fe_2As_2)_5$ with $T_c$ = 30 K. An anisotropic transient reflectivity induced by the fluctuations emerges below $T_{nem} \approx$ 120 K, coexisting and competing with the superconductivity ($\Delta_{SC}(0) \approx 11.5 \pm 0.1$ meV). Three coherent phonons were observed at 1.6 ($A_{1g}(1)$ mode), 3.5 ($E_g$ mode), and 4.7 THz ($A_{1g}(2)$ mode). The $A_{1g}$ phonons with $\lambda_{A_{1g}(2)} \approx 0.139 \pm 0.02$ deviate from the curves described by the phonon anharmonic effect at low temperatures. Our results provide critical information for understanding the relationship between the superconductivity and the nematic state in quasi-2D FeSCs and prove the $A_{1g}$ phonons are closely related to the superconductivity and nematic ordering in $(Ca_{0.85}La_{0.15})_{10}(Pt_3As_8)(Fe_2As_2)_5$.

This work was supported by the National Natural Science Foundation of China (Grants No. 92265101, No. 12074436, No. 11822411 and No.11961160699), the National Key Research and Development Program of China (Grants No. 2022YFA1604204 and No. 2018YFA0704200), J. Q. M would like to acknowledge support from the Science and Technology Innovation Program of Hunan Province (2022RC3068). H. Q. L was supported by the Youth Innovation Promotion Association of CAS (Grant No. Y202001). W. S. H was supported by the Postdoctoral Innovative Talent program (Grant No. BX2021018), and the China Postdoctoral Science Foundation (Grant No. 2021M700250).

* Corresponding author: hqluo@iphy.ac.cn
† Corresponding author: jqmeng@csu.edu.cn

[1] Y. Kamihara, T. Watanabe, M. Hirano, and H. Hosono, J. Am. Chem. Soc. **11**, 130 (2008).
[2] M. Hashimoto, I. M. Vishik, R. H. He, T. P. Devereaux, and Z. X. Shen, Nat. Phys. **10**, 483 (2014).
[3] D. C. Johnston, Adv. Phys. **59**, 803 (2010).
[4] J. Paglione and R. L. Greene, Nat. Phys. **6**, 645 (2010).
[5] H. H. Kuo, J. H. Chu, J. C. Palmstrom, S. A. Kivelson, and I. R. Fisher, Science **352**, 958 (2016).
[6] A. Dusza, A. Lucarelli, F. Pfuner, J. H. Chu, I. R. Fisher, and L. Degiorgi, Europhys. Lett. **93**, 37002 (2011).
[7] M. Yi, D. H. Lu, J. H. Chu, J. G. Analytis, A. P. Sorini, A. F. Kemper, B. Moritz, S. K. Mo, R. G. Moore, M. Hashimoto, W. S. Lee, Z. Hussain, T. P. Devereaux, I. R. Fisher, and Z. X. Shen, Pro. Natl. Acad. Sci. **108**, 6878 (2011).
[8] V. Hinkov, D. Haug, B. Fauqu, P. Bourges, Y. Sidis, A. Ivanov, C. Bernhard, C. T. Lin, and B. Keimer, Science **319**, 597 (2008).
[9] J. Wu, A. T. Bollinger, X. He, and I. Bozovic, Nature (London) **547**, 432 (2017).
[10] R. M. Fernandes, A. V. Chubukov, and J. Schmalian, Nat. Phys. **10**, 97 (2014).
[11] H. Yamase and R. Zeyher, Phys. Rev. B **88**, 180502(R) (2013).
[12] T. Agatsuma and H. Yamase, Phys. Rev. B **94**, 214505 (2016).
[13] M. D. Watson, T. K. Kim, A. A. Haghighirad, N. R. Davies, A. McCollam, A. Narayanan, S. F. Blake, Y. L. Chen, S. Ghannadzadeh, A. J. Schofield, M. Hoesch, C. Meingast, T. Wolf, and A. I. Coldea, Phys. Rev. B **91**, 155106 (2015).
[14] S. H. Baek, D. V. Efremov, J. M. Ok, J. S. Kim, J. van den Brink, and B. Buchner, Nat. Mater. **14**, 210 (2015).
[15] Q. Zhang, R. M. Fernandes, J. Lamsal, J. Yan, S. Chi, G. S. Tucker, D. K. Pratt, J. W. Lynn, R. W. McCallum, P. C. Canfield, T. A. Lograsso, A. I. Goldman, D. Vaknin, and R. J. McQueeney, Phys. Rev. Lett. **114**, 057001 (2015).
[16] M. G. Kim, R. M. Fernandes, A. Kreyssig, J. W. Kim, A. Thaler, S. L. Budko, P. C. Canfield, R. J. McQueeney, J. Schmalian, and A. I. Goldman, Phys. Rev. B **83**, 134522 (2011).
[17] C. W. Luo, P. C. Cheng, S. H. Wang, J. C. Chiang, J. Y. Lin, K. H. Wu, J. Y. Juang, D. A. Chareev, O. S. Volkova, and A. N. Vasiliev, npj Quantum Mater **2**, 32 (2017).
[18] S. H. Liu, C. F. Zhang, Q. Deng, H. H. Wen, J. X. Li, E. E. M. Chia, X. Y. Wang, and M. Xiao, Phys. Rev. B **97**, 020505(R) (2018).
[19] A. Patz, T. Li, S. Ran, R. M. Fernandes, J. Schmalian, S. L. Bud'ko, P. C. Canfield, I. E. Perakis, and J. Wang, Nat. Commun. **5**, 3229 (2014).
[20] L. Stojchevska, T. Mertelj, Jiun-Haw Chu, Ian R. Fisher, and D. Mihailovic, Phys. Rev. B **86**, 024519 (2012).
[21] V. K. Thorsmølle, M. Khodas, Z. P. Yin, C. L. Zhang, S. V. Carr, P. C. Dai, and G. Blumberg, Phys. Rev. B **93**, 054515 (2016).
[22] P. Massat, D. Farina, I. Paul, S. Karlsson, P. Strobel, P. Toulemonde, M. A. Measson, M. Cazayous, A. Sacuto, S. Kasahara, T. Shibauchi, Y. Matsuda, and Y. Gallais, Proc. Natl. Acad. Sci. **113**, 9177 (2016).
[23] W. Albrecht, T. Kruse, and H. Kurz, Phys. Rev. Lett. **69**, 1451 (1992).
[24] C. Zhang, Q. Y. Wu, W. S. Hong, H. Liu, S. X. Zhu, J. J. Song, Y. Z. Zhao, F. Y. Wu, Z. T. Liu, S. Y. Liu, Y. H. Yuan, H. Huang, J. He, S. L. Li, H. Y. Liu, Y. X. Duan, H. Q. Luo, and J. Q. Meng, Sci. China: Phys. Mech. Astron. **65**, 237411 (2022).
[25] S. X. Zhu, C. Zhang, Q. Y. Wu, X. F. Tang, H. Liu, Z. T. Liu, Y. Luo, J. J. Song, F. Y. Wu, Y. Z. Zhao, S. Y. Liu, T. Le, X. Lu, H. Ma, K. H. Liu, Y. H. Yuan, H. Huang, J. He, H. Y. Liu, Y. X. Duan, and J. Q. Meng, Phys. Rev. B **103**, 115108 (2021).
[26] S. Y. Liu, S. X. Zhu, Q. Y. Wu, C. Zhang, P. B. Song, Y. G. Shi, H. Liu, Z. T. Liu, J. J. Song, F. Y. Wu, Y. Z. Zhao, X. F. Tang, Y. H. Yuan, H. Huang, J. He, H. Y. Liu, Y. X. Duan, and J. Q. Meng, Results Phys **30**, 104816 (2021).
[27] T. Stürzer, G. Derondeau, and D. Johrendt, Phys. Rev. B **86**, 060516(R) (2012).
[28] N. Ni, J. M. Allred, B. C. Chan, and R. J. Cava, PNAS **108**, E1019 (2011).

[29] N. Ni, W. E. Straszheim, D. J. Williams, M. A. Tanatar, R. Prozorov, E. D. Bauer, F. Ronning, J. D. Thompson, and R. J. Cava, Phys. Rev. B **87**, 060507(R) (2013).
[30] P. W. Gao, L. L. Sun, N. Ni, J. Guo, Q. Wu, C. Zhang, D. C. Gu, K. Yang, A. G. Li, S. Jiang, R. J. Cava, and Z. X. Zhao, Adv. Mater. **26**, 2346 (2014).
[31] A. Sapkota, G. S. Tucker, M. Ramazanoglu, W. Tian, N. Ni, R. J. Cava, R. J. McQueeney, A. I. Goldman, and A. Kreyssig, Phys. Rev. B **90**, 100504(R) (2014).
[32] Y. I. Seo, W. J. Choi, S. I. Kimura, Y. Bang, and Y. S. Kwon, Phys. Rev. B **95**, 094510 (2017).
[33] Z. Z. Li, W. S. Hong, H. L. Zhou, X. Y. Ma, U. Stuhr, K. Y. Zeng, L. Ma, Y. Xiang, H. Yang, H. H. Wen, J. P. Hu, S. L. Li, and H. Q. Luo, unpublished (2022).
[34] See Supplemental Material for additional experiment results and suporting data analysis, which includes Refs. [35–37]
[35] D. Dvorsek, V. V. Kabanov, J. Demsar, S. M. Kazakov, J. Karpinski, and D. Mihailovic, Phys. Rev. B **66**, 020510(R) (2002).
[36] A. Baum, Ying Li, M. Tomić, N. Lazarević, D. Jost, F. Löffler, B. Muschler, T. Böhm, J. H. Chu, I. R. Fisher, R. Valentí, I. I. Mazin, and R. Hackl, Phys. Rev. B **98**, 075113 (2018).
[37] Q. Wu, H. X. Zhou, Y. L. Wu, L. L. Hu, S. L. Ni, Y. C. Tian, F. Sun, F. Zhou, X. L. Dong, Z. X. Zhao, and J. M. Zhao, Chin. Phys. Lett. **37**, 097802 (2020).
[38] K. H. Lin, K. J. Wang, C. C. Chang, Y. C. Wen, D. H. Tsai, Y. R. Wu, Y. T. Hsieh, M. J. Wang, B. Lv, Paul Ching-Wu Chu, and M. K. Wu, Phys. Rev. B **90**, 174502 (2014).
[39] A. Rothwarf and B. N. Taylor, Phys. Rev. Lett. **19**, 27 (1967).
[40] Y. H. Liu, Y. Toda, K. Shimatake, N. Momono, M. Oda, and M. Ido, Phys. Rev. Lett. **101**, 137003 (2008).
[41] E. E. M. Chia, D. Talbayev, J. X. Zhu, H. Q. Yuan, T. Park, J. D. Thompson, C. Panagopoulos, G. F. Chen, J. L. Luo, N. L. Wang, and A. J. Taylor, Phys. Rev. Lett. **104**, 027003 (2010).
[42] J. Qi, T. Durakiewicz, S. A. Trugman, J. X. Zhu, P. S. Riseborough, R. Baumbach, E. D. Bauer, K. Gofryk, J. Q. Meng, J. J. Joyce, A. J. Taylor, and R. P. Prasankumar, Phys. Rev. Lett. **111**, 057402 (2013).
[43] Y. Z. Zhao, Q. Y. Wu, C. Zhang, B. Chen, W. Xia, J. J. Song, Y. H. Yuan, H. Liu, F. Y. Wu, X. Q. Ye, H. Y. Zhang, H. Huang, H. Y. Liu, Y. X. Duan, Y. F. Guo, J. He, and J. Q. Meng, arXiv:2209.10169.
[44] G. P. Serge, N. Gedik, J. Orenstein, D. A. Bonn, Ruixing Liang, and W. N. Hardy, Phys. Rev. lett. **88**, 137001 (2002).
[45] M. A. Surmach, F. Brückner, S. Kamusella, R. Sarkar, P. Y. Portnichenko, J. T. Park, G. Ghambashidze, H. Luetkens, P. K. Biswas, W. J. Choi, Y. I. Seo, Y. S. Kwon, H. H. Klauss, and D. S. Inosov, Phys. Rev. B **91**, 104515 (2015).
[46] E. Thewalt, I. M. Hayes, J. P. Hinton, A. Little, S. Patankar, L. Wu, T. Helm, C. V. Stan, N. Tamura, J. G. Analytis, and J. Orenstein, Phys. Rev. Lett. **121**, 027001 (2018).
[47] K. Y. Choi, D. Wulferding, P. Lemmens, N. Ni, S. L. Bud'Ko, and P. C. Canfield, Phys. Rev. B **78**, 212503 (2008).
[48] R. A. Jishi and H. M. Alyahyaei, Adv. Condens. Matter Phys. **2010**, 804343 (2010).
[49] R. Merlin, Solid State Commun **102**, 207 (1997).
[50] A. P. Litvinchuk, B. Lv, and C. W. Chu, Phys. Rev. B **84**, 092504 (2011).
[51] J. Menéndez and M. Cardona, Phys. Rev. B **29**, 2051 (1984).
[52] S. Y. Zhang, J. Q. Guan, X. Jia, B. Liu, W. H. Wang, F. S. Li, L. L. Wang, X. C. Ma, Q. K. Xue, J. D. Zhang, E. W. Plummer, X. T. Zhu, and J. D. Guo, Phys. Rev. B **94**, 081116(R) (2016).

Supplementary material

# The Hidden Nematic Fluctuation in the Triclinic $Ca_{8.5}La_{1.5}(Pt_3As_8)(Fe_2As_2)_5$ Superconductor Revealed Ultrafast Optical Spectroscopy

Qi-Yi Wu,[1] Chen Zhang,[1] Ze-Zhong Li,[2] Wen-Shan Hong,[2,3] Hao Liu,[1] Jiao-Jiao Song,[1] Yin-Zou Zhao,[1] Ya-Hua Yuan,[1] Bo Chen,[1] Xue-Qing Ye,[1] Shiliang Li,[2,4] Jun He,[1] H. Y. Liu,[5] Yu-Xia Duan,[1] Hui-Qian Luo,[2,4,*] and Jian-Qiao Meng[1,†]

[1]*School of Physics and Electronics, Central South University, Changsha 410083, Hunan, China*
[2]*Beijing National Laboratory for Condensed Matter Physics, Institute of Physics, Chinese Academy of Sciences, Beijing 100190, China*
[3]*International Center for Quantum Materials, School of Physics, Peking University, Beijing 100871, China*
[4]*Songshan Lake Materials Laboratory, Dongguan 523808, China*
[5]*Beijing Academy of Quantum Information Sciences, Beijing 100193, China*

E-Mail：hqluo@iphy.ac.cn
jqmeng@csu.edu.cn

**The supplemental materials to ‘The hidden nematic fluctuation in the Triclinic $Ca_{8.5}La_{1.5}(Pt_3As_8)(Fe_2As_2)_5$ Superconductor Revealed Ultrafast Optical Spectroscopy’ contains ‘Temperature dependence of resistance and magnetic susceptibility’, ‘Tri-exponential fitting results along $\phi = 0°$ and $\phi = 90°$’, ‘Polarization dependence of the superconducting response’, ‘Linear fluence dependence’, ‘Weak fluence detects superconducting component’, ‘Coherent phonons along $\phi = 0°$’, and ‘Calculation of electron-phonon coupling constant’.**

## 1. Temperature dependence of resistance and magnetic susceptibility

Figure S1 displays the temperature dependence of in-plane resistance and magnetic susceptibility of $Ca_{8.5}La_{1.5}(Pt_3As_8)(Fe_2As_2)_5$ measured on the PPMS. The drop of resistance and magnetization occurs around 30 K, which means the superconducting temperature of $Ca_{8.5}La_{1.5}(Pt_3As_8)(Fe_2As_2)_5$ is around $T_c = 30 \pm 3$ K.

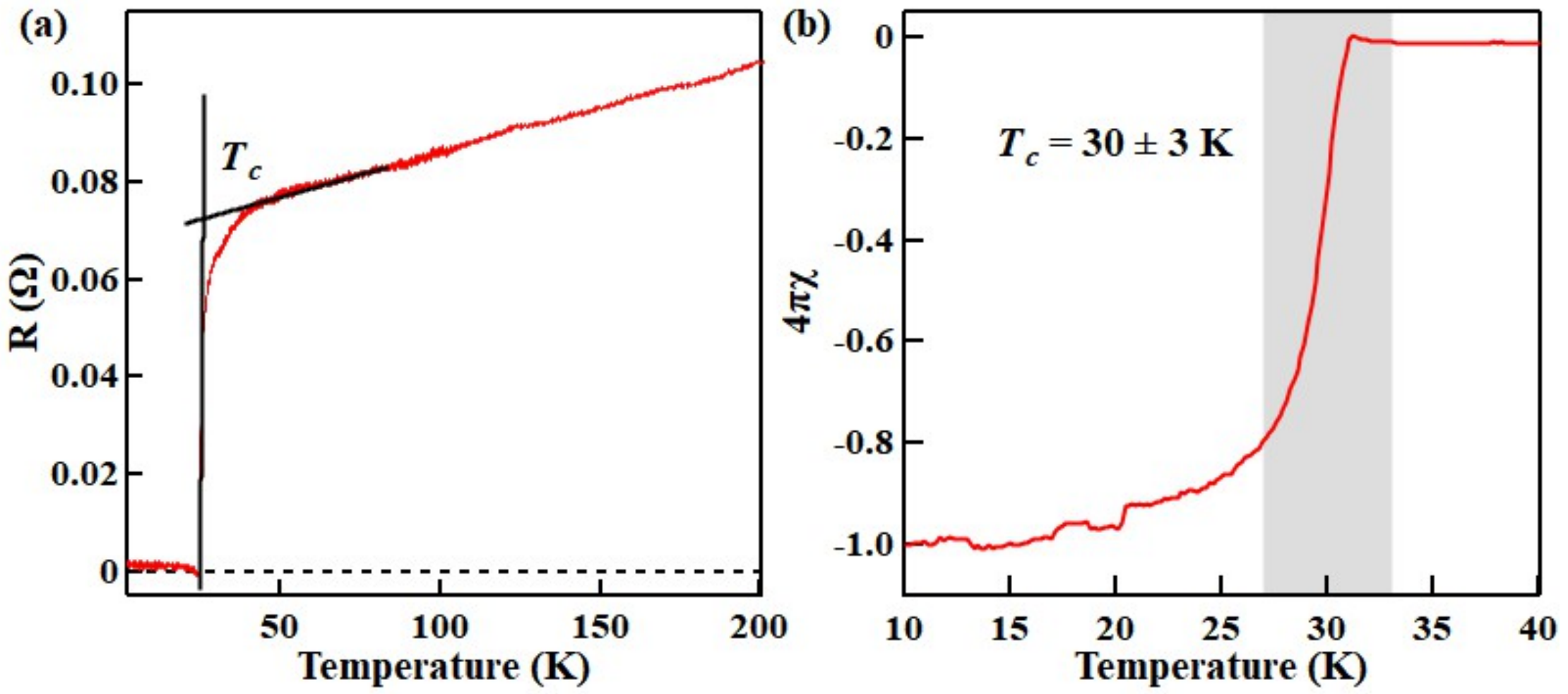

**Fig. S1** The temperature dependence of resistance (a) and magnetic susceptibility (b) for $Ca_{8.5}La_{1.5}(Pt_3As_8)(Fe_2As_2)_5$.

## 2. Tri-exponential fitting results along $\phi = 0^\circ$ and $\phi = 90^\circ$

Figure S2 displays the original data and the fitted curves using tri-exponential decays from 4 to 180K at a pump fluence of ~3.8 μJ/cm$^2$, along $\phi = 0^\circ$ (a) and $\phi = 90^\circ$ (b). The fitted curves agree well with the original data, indicating that using the tri-exponential decays to fit the data is reasonable.

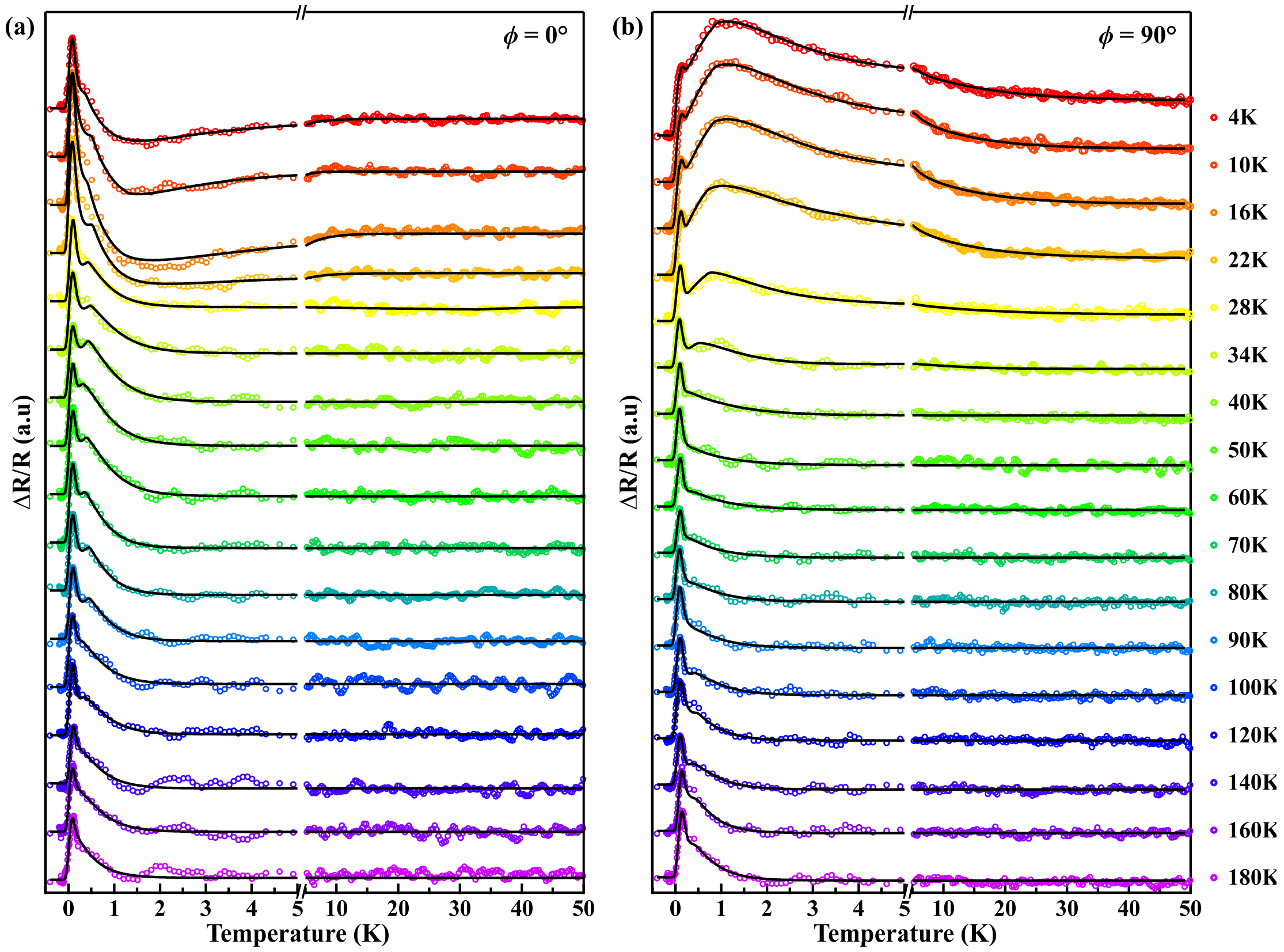

**Fig. S2** The original data (open circles) and the fitting results (solid lines) from 4 to 180 K, along $\phi = 0^\circ$ (a) and $\phi = 90^\circ$ (b).

## 3. Polarization dependence of the superconducting response

Figure S3 displays the angular dependence of amplitude $A_3$ at 5 K extracted by Eq. (1) in the main text. $A_3$ has a *d*-wave shape (Fig. S3(a)), and the sign change when the polarization angle rotates from $\phi = 0°$ to $\phi = 90°$. According to Dvorsek et al.'s [1] methodology, the polarization dependence of $A_3$ can be explained by the anisotropy of the probe transition matrix elementary. In the pump-probe experiment under low pump fluence, the reflectivity $\Delta R$ is proportional to the photoinduced absorption $\Delta\alpha$ and mainly contributed by the inter-band resonance transition, i.e.:

$$\Delta R \propto \Delta\alpha \propto \int d\epsilon N(\epsilon) \left| M(\epsilon,\omega) \right|^2 [f'(\epsilon) - f(\epsilon)] \qquad (1)$$

where $N(\epsilon)$ is the density of the electronic states, $f'(\epsilon)$ is the nonequilibrium distribution function of the charge carriers, $f(\epsilon)$ is the equilibrium distribution function of the charge carriers, and $M(\epsilon, \omega)$ is the dipole matrix element for the transition. The

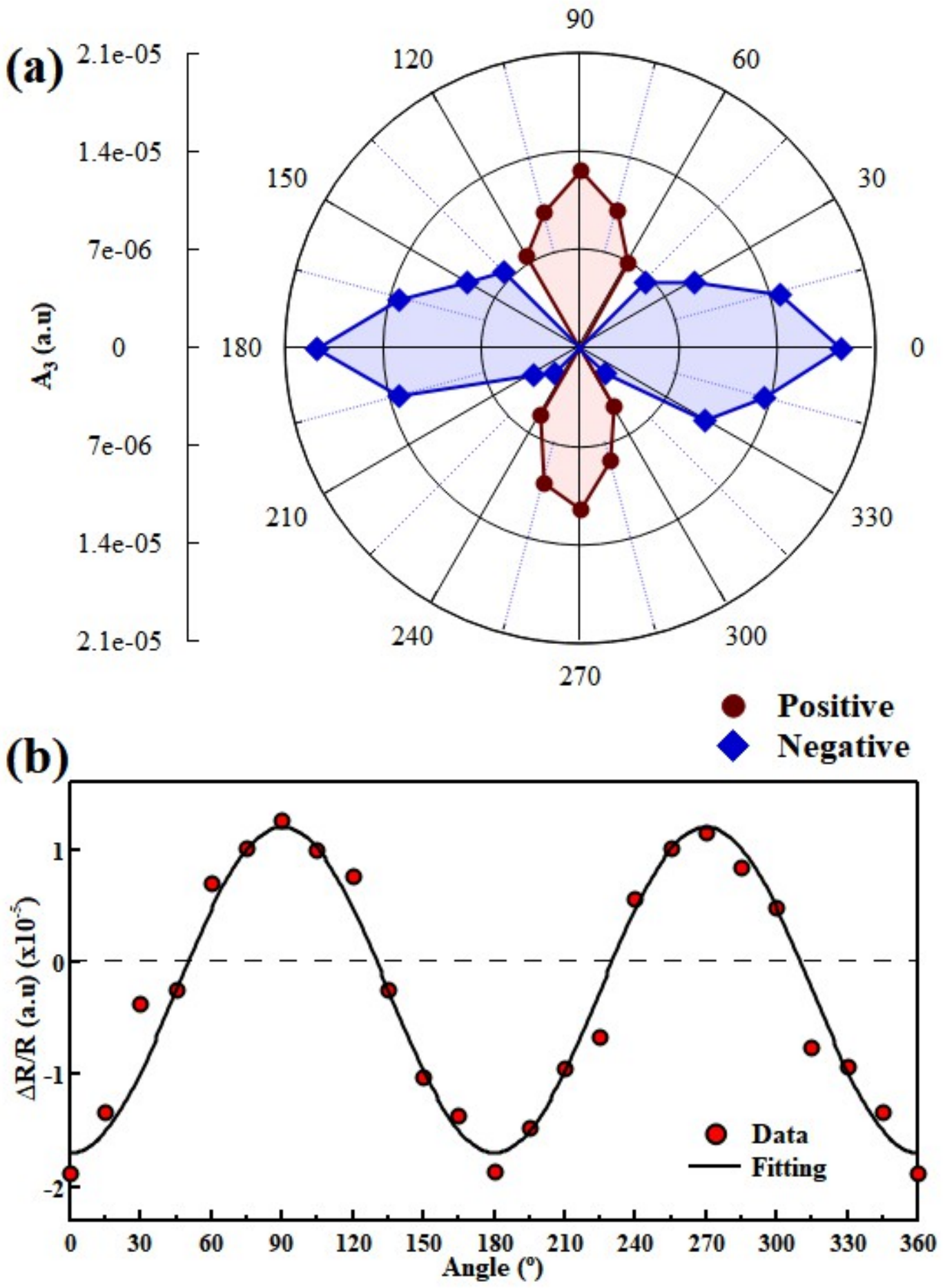

**Fig. S3** (a) The polar graph of the SC component. (b) The normalized reflectivity as a function of the polarization angle, the dotted line is the Fitting curve by using Eq. (2).

integral is taken in the vicinity of the Fermi energy $\epsilon_F$.

If the $M(\epsilon, \omega)$ is a constant over the energy range of 1.55eV< $\epsilon$- $\epsilon_F$<1.55eV, the ΔR = Δα = 0 because of the conservation of particles. Thus, the $M(\epsilon, \omega)$ is not a constant in the experiment. The expression of the probe absorption with polarization dependence can be written as:

$$\Delta R \propto \Delta\alpha \propto M_0[\gamma_x \sin^2(\theta) + \gamma_y \cos^2(\theta)]\Delta n \quad (2)$$

where the $\theta$ is the polarization angle, the $\gamma_{x,y}$ is the derivation of dipole matrix element at $\epsilon_F$, i.e., $\gamma_{x,y} = dM_{x,y}/d\epsilon$, and $\Delta n$ is the number of the photoexcited quasiparticles. The $\gamma_{x,y}$ can have different values and signs. According to Eq. (2), the polarization dependence of $A_3$ can be well described using $\gamma_x$= 7.4 and $\gamma_y$= -10.4, as shown by the solid black line in Fig. S3(b).

## 4. Linear fluence dependence

The initial amplitudes at various temperatures increase linearly with the pump fluence, as shown in Fig. S4. This linear relationship indicates that the photo-excited QPs densities are proportional to the pump fluence. Therefore, the fastest component characterized by $A_1$ and $\tau_1$ is the *e*-*e* scattering process.

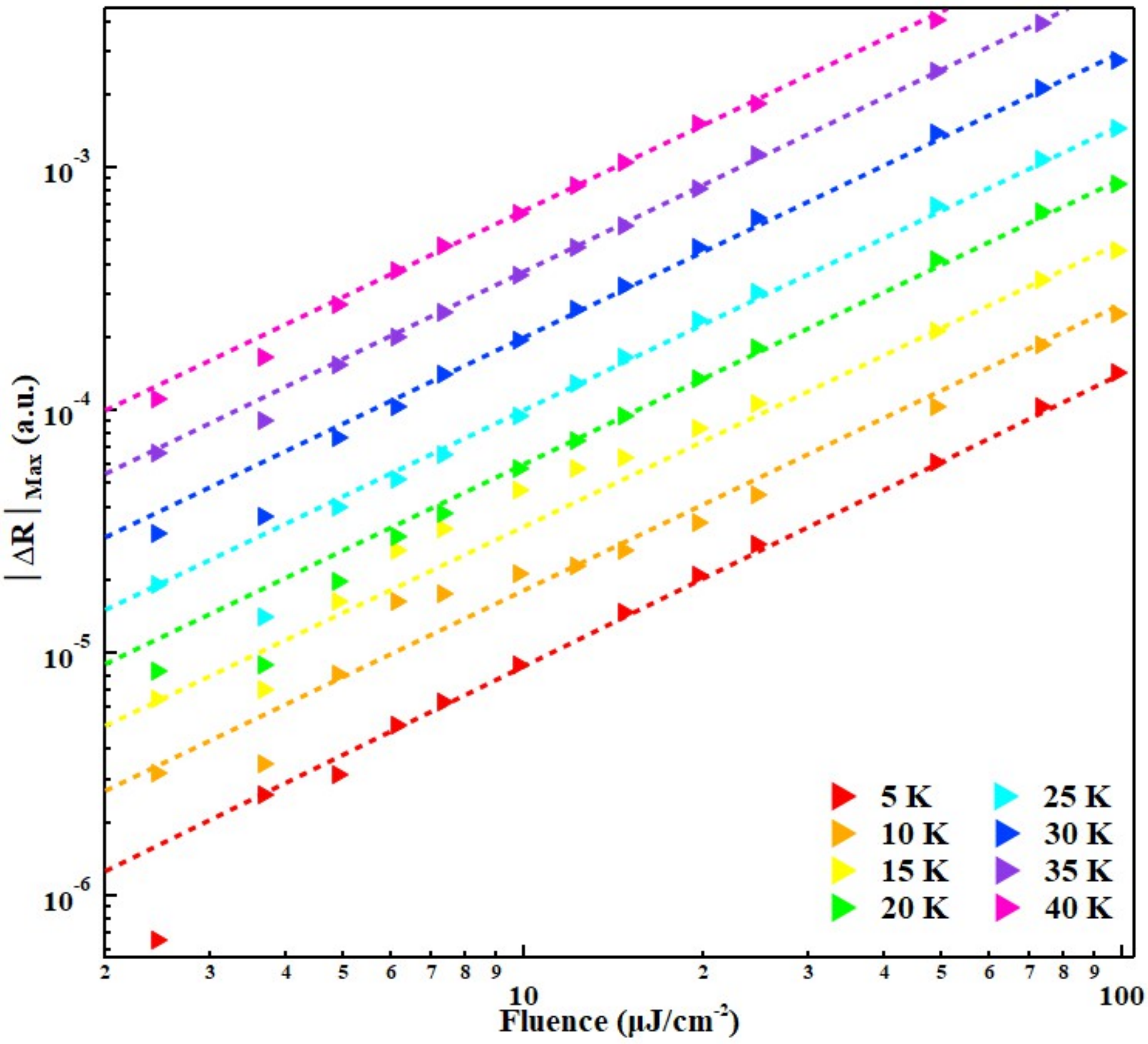

**Fig. S4** Initial amplitude $|\Delta R|_{max}$ as a function of pump fluence at various temperatures (offset for clarity). Dashed lines show linear fits to data with a slope of 1.

## 5. Weak fluence detects superconducting component

Choosing an appropriate pump fluence to perform the temperature-dependent measurement is necessary to avoid destroying the superconducting component. Fig. S5 (a) displays the normalized transient reflectivity at 10 K for various pump fluences. The normalized $\Delta R/R$ shows no change at low fluence. However, the superconducting component decreases with the increasing fluence when the pump fluence exceeds the critical fluence. The amplitude $|A_{SC}|$ and relaxation time of the superconducting component as a function of fluence is obtained by using Eq. (1) in the main text shown in Fig. S5(b) and (c). The amplitude increases as the fluence increase, while the relaxation time is almost a constant below ~ 7 μJ/cm$^2$. When the pump fluence exceeds 7 μJ/cm$^2$, the $|A_{SC}|$ appears a saturated behavior, and the $\tau_{SC}$ increases as fluence. Thus the 7 μJ/cm$^2$ is defined as the saturated fluence without destructing the SC component. The data shown in the main text was measured at ~3.8 μJ/cm$^2$ lower than the threshold.

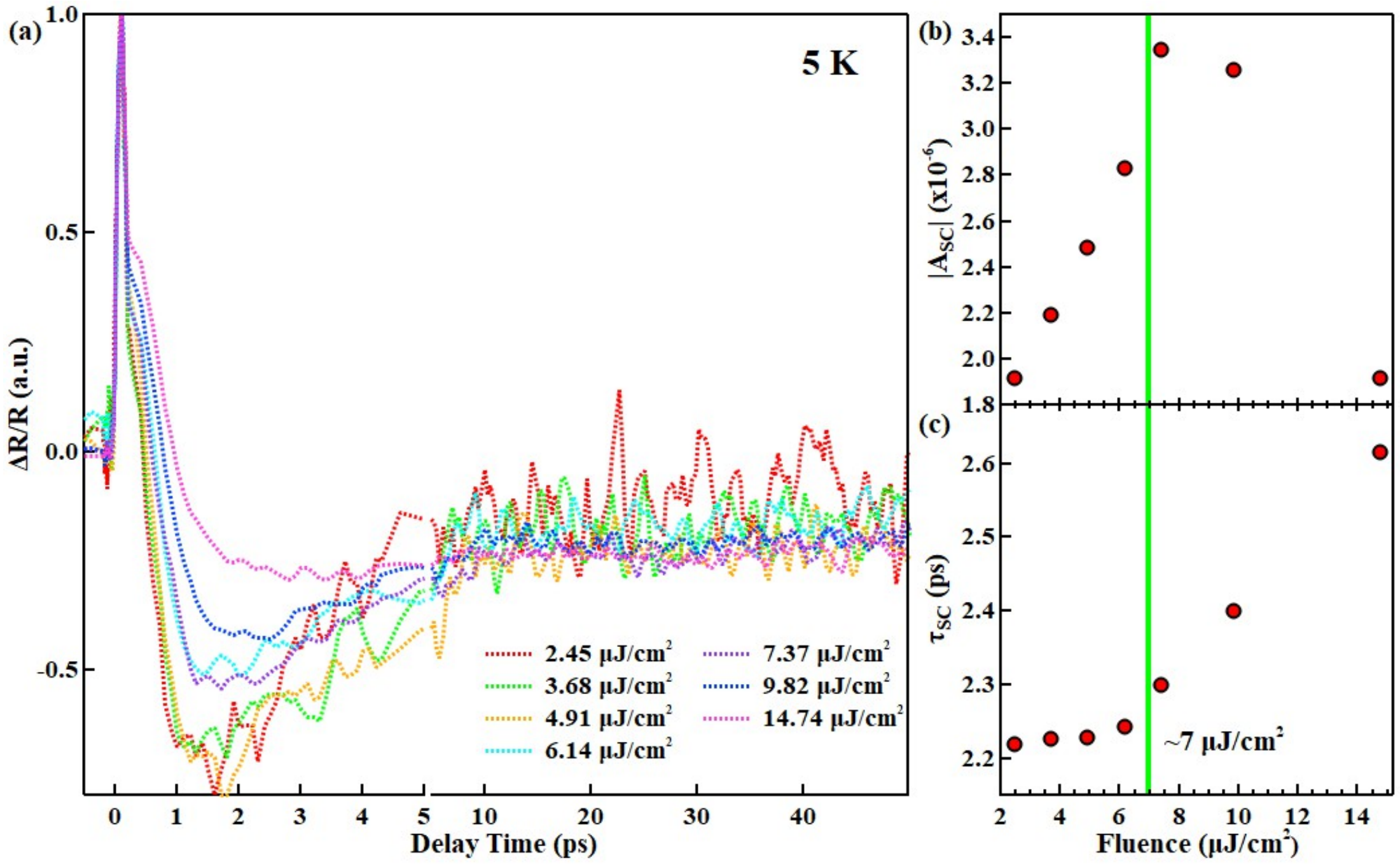

**Fig. S5** (a) The normalized transient reflectivity at 10 K for various pump fluence. (b)**,** (c) The amplitude $|A_{SC}|$ and relaxation time $\tau_{SC}$ of the SC component as a function of pump fluence.

## 6. Coherent phonons along $\phi = 0^{\circ}$

Figure S6 displays the 2D map of FFT results as a function of temperature and frequency along $\phi = 0^{\circ}$ and $\phi = 90^{\circ}$. We note that the splitting peaks from $\omega_2$ mode with a frequency around 3.1 THz are visible at all measured temperatures from the FFT result as a function of temperature and frequency along $\phi = 90^{\circ}$ [Fig. S6(a)], while it is hard to distinguish from the background for $\phi = 0^{\circ}$ [Fig. S6(b)]. The split persisting to room temperature can account for the splitting of $E_g$ phonon into $B_{2g}$ and $B_{3g}$ modes [2,3] resulting from the triclinic crystal structure of CaLa10-3-8 might affect the tetragonal structure of FeAs layers. Moreover, since the $B_{2g}(1)$ and $B_{3g}(1)$ modes have different sensitivities to the laser polarization, it is difficult to distinguish this splitting along $\phi = 0^{\circ}$. The temperature dependence of extracted $\omega_{2,0}$ [Fig. S6(e)]and $\omega_{2,90}$ [Fig. 3(e)] can be described by the anharmonic effect and exhibits no anomalies around $T_c$, indicating the $E_g$ mode might be independent of superconductivity. Moreover, in contrast to the $A_{1g}$ modes behaviors described in the main text, the temperature dependence of $A_{1g}$ modes along $\phi = 0^{\circ}$ match the curves obtained by the phonon anharmonic effect, giving another evidence that there is strong anisotropy in CaLa10-3-8.

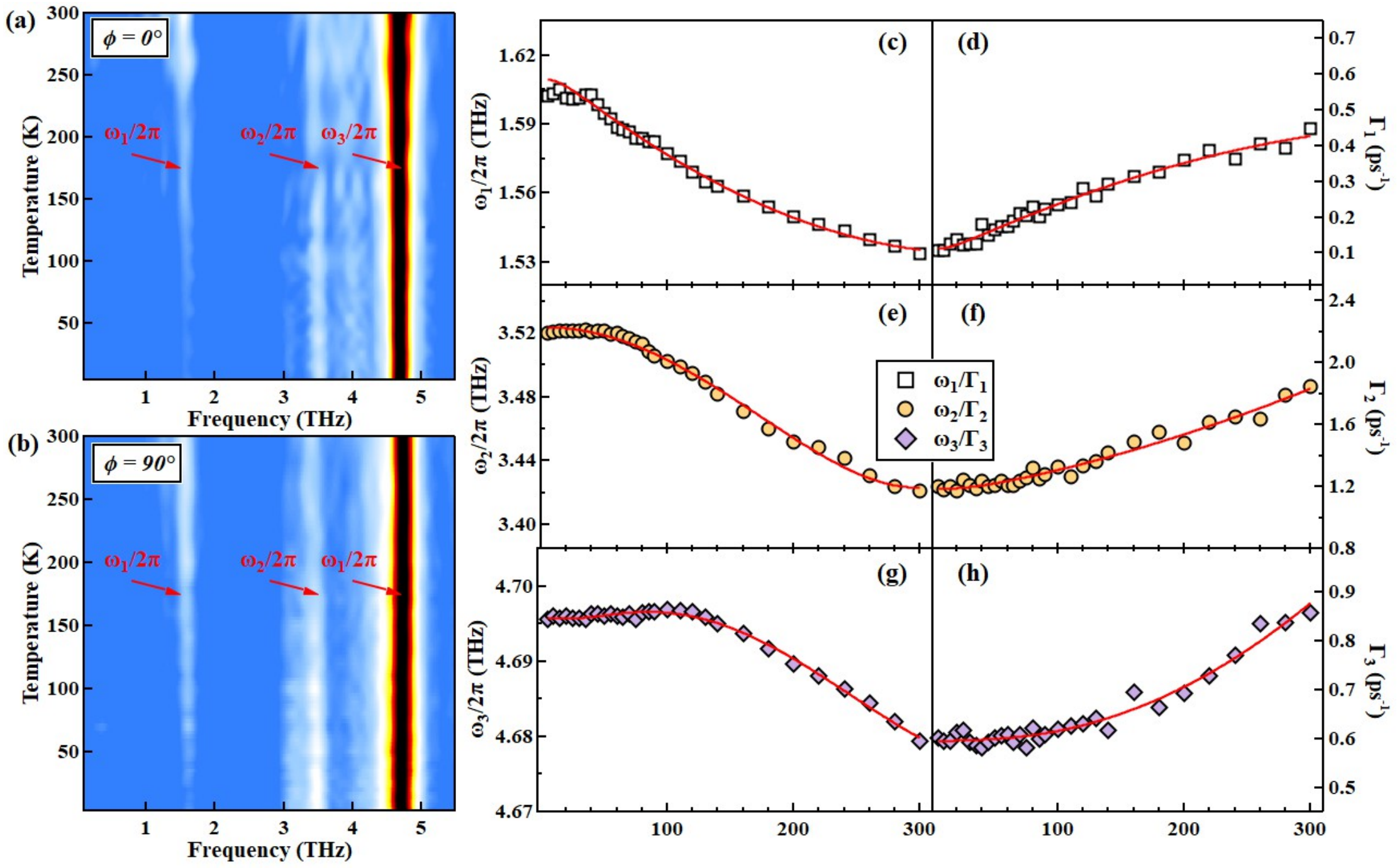

**Fig. S6** (a) and (b) 2D pseudocolor mapping of FFT results as a function of temperature and frequency along $\phi = 0^{\circ}$ and $\phi = 90^{\circ}$, respectively. (c)-(h) The frequency and damping rate of three phonon modes along $\phi = 0^{\circ}$. The solid lines are anharmonic effect fitting curves.

## 7. Calculation of electron-phonon coupling constant

In metal, one of the relaxation processes under high fluence is the electron-phonon scattering process. Fig. S7 shows the temperature-dependence of the electron-phonon scattering time $\tau_{e\text{-}ph}$ along two directions. For both polarizations, the $\tau_{e\text{-}ph}$ decreases as the cooling temperature and remains constant at a range of temperatures, with a divergent behavior at low temperatures. However, the details of the temperature-dependent curves are different. For $\phi = 0°$, the slope of $\tau_{e\text{-}ph}$ at low temperatures is steeper, and the flat range is wider. Above Debye temperature (257 K [4]), the $\tau_{e\text{-}ph}$ along both polarizations present a linear relationship with the temperature and can be fit well by the EMTM model (the dotted line in (a) and (b)). Thus, the second moment of the Eliashberg-function can be deduced by the equation $\tau_{e\text{-}ph} = 2\pi k_B T / 3\hbar\lambda\langle\omega^2\rangle$ [5], i.e., $\lambda\langle\omega^2\rangle \approx 1.22 \times 10^{26}$ Hz$^2$ (~ 52 meV$^2$). Considering the phonon frequency used to calculate the $\lambda_{A1g}$ in other iron-based superconductors, we use the $\omega_3/2\pi$ mode with the frequency of 4.7 THz (~ 19.4 meV) to calculate the $\lambda_{A1g}$ of Ca10-3-8. The *e-ph* coupling constant is estimated to be $\lambda_{A1g(2)} \approx 0.14 \pm 0.02$ for both polarizations.

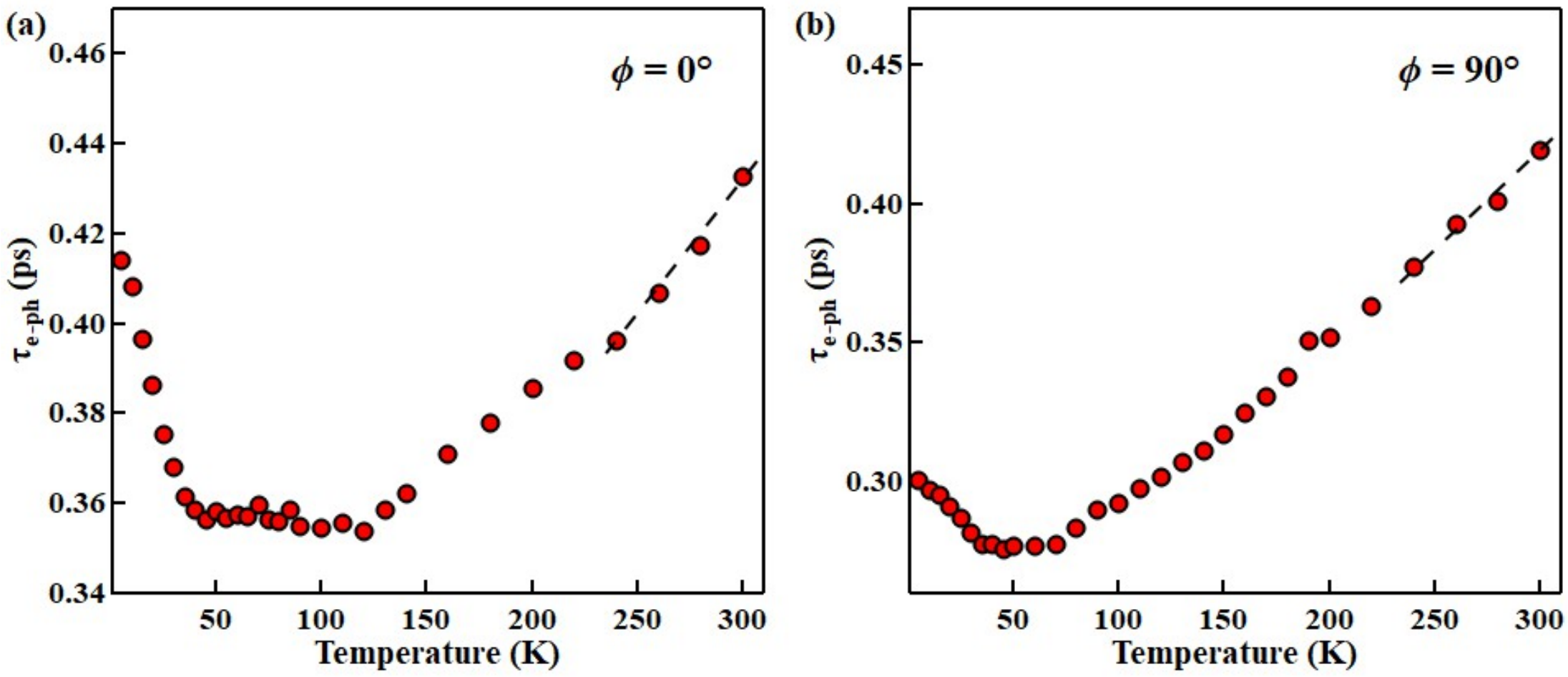

**Fig. S7** The relaxation time of *e-ph* scattering process along $\phi = 0°$ (a) and $\phi = 90°$ (b).

# Reference

[1]. D. Dvorsek, V. V. Kabanov, J. Demsar, S. M. Kazakov, J. Karpinski, and D. Mihailovic, Phys. Rev. B **66**, 020510(R) (2002).

[2]. Baum, Ying Li, M. Tomić, N. Lazarević, D. Jost, F. Lö ffler, B. Muschler, T. Bö hm, J. H. Chu, I. R. Fisher, R. Valentí, I. I. Mazin, and R. Hackl, Phys. Rev. B **98**, 075113 (2018).

[3]. Q. Wu, H. X. Zhou, Y. L. Wu, L. L. Hu, S. L. Ni, Y. C. Tian, F. Sun, F. Zhou, X. L. Dong, Z. X. Zhao, and J. M. Zhao, Chin. Phys. Lett. **37**, 097802 (2020).

[4]. N. Ni, W. E. Straszheim, D. J. William, M. A. Tanatar, R. Prozorov, E. D. Bauer, F. Ronning, J. D. Thompson, and R. J. Cava, Phys. Rev. B **87** 060507(R) (2013).

[5]. L. Stojchevska, T. Mertelj, J.-H. Chu, I. R. Fisher, and D. Mihailovic, Phys. Rev. B **86**, 024519 (2012).